\documentclass[conference]{IEEEtran}
\IEEEoverridecommandlockouts
\usepackage{cite}
\usepackage{amsmath,amssymb,amsfonts}
\usepackage{algorithmic}
\usepackage{graphicx}
\usepackage{textcomp}
\usepackage{xcolor}
\usepackage{url}
\usepackage{tabularx}
\def\BibTeX{{\rm B\kern-.05em{\sc i\kern-.025em b}\kern-.08em
    T\kern-.1667em\lower.7ex\hbox{E}\kern-.125emX}}
\newboolean{blind}
\setboolean{blind}{false}
\begin{document}

\title{Accelerating netty-based applications through transparent InfiniBand support}

\ifblind
\author{
	Project title and authors have been omitted for double blind reviews.\\
	\ \\
	\ \\
	\ \\
	\ \\
	\ \\
}
\else
\author{
	\IEEEauthorblockN{Fabian Ruhland}
	\IEEEauthorblockA{
		\textit{Department Operating Systems} \\
		\textit{Heinrich Heine University}\\
		Düsseldorf, Germany \\
		fabian.ruhland@hhu.de 
	}
	
	\and
	
	\IEEEauthorblockN{Filip Krakowski}
	\IEEEauthorblockA{
		\textit{Department Operating Systems} \\
		\textit{Heinrich Heine University}\\
		Düsseldorf, Germany \\
		filip.krakowski@hhu.de
	}
	
	\and
	
	\IEEEauthorblockN{Michael Schöttner}
	\IEEEauthorblockA{
		\textit{Department Operating Systems} \\
		\textit{Heinrich Heine University}\\
		Düsseldorf, Germany \\
		michael.schoettner@hhu.de
	}
}
\fi

\maketitle

\begin{abstract}
	Many big-data frameworks are written in Java, e.g. Apache Spark, Flink and Cassandra. These systems use the networking framework \emph{netty} which is based on Java NIO. While this allows for fast networking on traditional Ethernet networks, it cannot fully exploit the whole performance of modern interconnects, like InfiniBand, providing bandwidths of 100 Gbit/s and more.\\
	In this paper we propose netty support for \emph{hadroNIO}, a Java library, providing transparent InfiniBand support for Java applications based on NIO. hadroNIO is based on \emph{UCX}, which supports several interconnects, including InfiniBand. We present hadroNIO extensions and optimizations for supporting netty. The evaluations with microbenchmarks, covering single- and multi-threaded scenarios, show that it is possible for netty applications to reach round-trip times as low as 5 \textmu s and fully utilize the 100 Gbit/s bandwidth of high-speed NICs, without changing the application's source code. We also compare hadroNIO with traditional sockets, as well as libvma and the results show, that hadroNIO offers a substantial improvement over plain sockets and can outperform libvma in several scenarios.
\end{abstract}

\begin{IEEEkeywords}
	High-speed networks, Cloud computing, Ethernet, InfiniBand, OpenUCX, Java
\end{IEEEkeywords}
\section{Introduction}
\label{introduction}

Modern big-data applications need to operate on large data sets, often using well-known big-data frameworks, such as Apache Spark\cite{spark}, Flink\cite{flink} or Cassandra\cite{cassandra}. Many of these systems are written in Java, relying on Java NIO. Java NIO provides developers with the tools for building large-scale networking applications, by allowing a single thread to handle multiple connections asynchronously, thus being able to scale with the amount of CPU cores available in a system.

However, its API has a steep learning curve compared to traditional Java sockets, thread management is still being left to the programmer and buffers need to be allocated manually, requiring a sophisticated buffer management to prevent performance penalties by repeated allocations. Thus, many applications do not use Java NIO directly, but are based on \emph{netty}, an asynchronous event-driven network application framework\cite{netty}. It abstracts the complexity introduced by Java NIO, implements buffer pooling based on reference counting, and automatically uses as many worker threads, as there are CPU cores available. It is also highly configurable, rendering it a powerful, yet easy-to-use networking library.

Netty is widely adopted in the Java community as the standard framework for fast and scalable networking and is used in many projects, e.g. Apache BookKeeper\cite{bookkeeper} or Ratis\cite{ratis}, which implements the Raft\cite{raft} algorithm in Java. Additionally, it serves as the base for other networking libraries, like the widely used RPC framework gRPC\cite{grpc}, as well as many more projects\cite{netty-related}. Its relevance is further underlined by the amount of organizations, that incorporate netty into their projects, such as Google, Facebook and IBM\cite{netty-adopters}.

However, since netty is based on Java NIO, which relies on traditional sockets, it cannot use the full potential of modern network interconnects, like InfiniBand or high-speed Ethernet. While the socket API is compatible with high-speed Ethernet NICs and can be used with InfiniBand cards via the kernel module \emph{IP over InfiniBand}\cite{ipoib}, it uses the kernel's network stack, involving context switches between user and kernel space, for exchanging network data, thus imposing a high performance penalty, especially regarding latency.

This problem has been addressed in the past, with different native and Java-based solutions, which came in form of user space TCP-stacks, transparent libraries offloading traffic to high-speed NICs or kernel modules, replacing the traditional TCP implementation. However, many of these solutions are not supported anymore and introduce their own sets of problems, which we discuss in Section \ref{related-work}.

We proposed \emph{hadroNIO} in 2021\cite{hadroNIO}, a Java library, which transparently replaces the default NIO implementation, offloading traffic via the \emph{Unified Communication X} framework (UCX)\cite{ucx}. UCX is a native library, providing a unified API for multiple transport types (including InfiniBand) and offering a multitude of communication models, such as streaming, tagged messaging, active messaging and RDMA. It automatically detects all available transports and chooses the fastest one, but can also be configured to use a specific NIC or utilize multiple interconnects in a multi-rail setup. It officially supports Java via a JNI-based binding called JUCX. We already have shown that hadroNIO provides huge performance improvements over using traditional sockets in a single-connection setup, using a microbenchmark based directly on Java NIO\cite{hadroNIO}.

In this paper, we present the extensions and optimizations, introduced in hadroNIO and evaluate its performance with netty-based microbenchmarks using multiple connections on high-speed networking hardware, capable of 100 Gbit/s bandwidth.

The contributions of this paper are:
\begin{itemize}
	\item An overview of existing netty-compatible acceleration approaches
	\item Design and implementation of hadroNIO extensions for supporting netty
	\item Evaluations using microbenchmarks on 100 GBit/s hardware, showing the benefits of the proposed solution
\end{itemize}

The paper is structured as follows: Section \ref{related-work} presents related work, discussing alternative acceleration solutions. Section \ref{hadronio-changes} elaborates on updates to hadroNIO, followed by Section \ref{benchmark-architecture}, which presents the architecture of our microbenchmarks. Evaluation results are discussed in Section \ref{evaluation}, while Section \ref{conclusion} concludes this paper and provides ideas for future work.
\section{Related work}
\label{related-work}

Modern high-speed NICs from Mellanox can be configured to use either InfiniBand or Ethernet as link layer protocol. Choosing Ethernet makes these cards fully compatible with the standard socket API, while still being programmable via the ibverbs library. Regardless of the link layer protocol, traditional sockets do not suffice for using the full potential of such a NIC.

While we are not aware of any alternative NIO implementations, there are several solutions for accelerating traditional sockets, with only few being still actively maintained. Typically, these can come in three different shapes: kernel modules, native libraries and Java libraries. Since the default NIO implementation is based on classic sockets, these solutions can be used to accelerate Java NIO applications. We have already evaluated some of these solutions, using socket-based microbenchmarks\cite{observatory} and compared them to hadroNIO with another microbenchmark, directly using the NIO API\cite{hadroNIO}.

\subsection{Kernel modules}
\textbf{IP over InfiniBand}\cite{ipoib} exposes InfiniBand devices as standard network interfaces, enabling applications to use them by simply binding to an IP address, associated with such a device. This solution does not require any preloading of libraries, making it the easiest to use. However, it relies on the kernel's network stack, thus requiring context switches which impose a large performance overhead, rendering it unattractive for applications requiring low latency.

\textbf{Fastsocket}\cite{fastsocket} replaces the Linux kernel's TCP implementation, aiming to provide better scaling with multiple CPU cores. It has been evaluated using up to 24 cores using 10 Gbit/s Ethernet NICs, showing much better scalability than the default TCP implementation. Fastsocket consists of kernel level optimizations, a kernel module and a user space library. It requires a custom kernel, based on Linux 2.6.32 and officially only supports CentOS 6.5, which is outdated by now. While it would be interesting to see how such an integrated solution would perform on modern high-speed Ethernet hardware, it does not seem to be in active development anymore.

\subsection{Native libraries}
\textbf{mTCP}\cite{mtcp} is a TCP-stack, running completely in user space. As Fastsocket, it primarily aims at high scalability, which it achieves by being independent from the kernel's network stack, alleviating the need for context switches in network applications. Contrary to the other solutions, it is not transparent and requires rewriting parts of an application's network code. It has no official support for Java, but there is an unofficial binding called JmTCP, based on the Java Native Interface (JNI). However, it does not seem to be actively maintained, probably requiring Java applications to manually access mTCP via JNI or the experimental Foreign Function \& Memory API (Project Panama)\cite{panama}. Since it is neither transparent, nor officially supports Java, mTCP does not fit our use case of accelerating netty-based applications.

\textbf{libvma}\cite{libvma} is a library developed in C/C++ by Mellanox, transparently offloading socket traffic to high-speed Ethernet or InfiniBand NICs. It can be preloaded to any socket-based application (using \emph{LD\_PRELOAD}), enabling full kernel bypass without the need to modify an application's code. However, libvma requires the \emph{CAP\_NET\_RAW} capability, which might not be available, depending on the cluster environment.

While it is highly configurable by exposing many parameters, allowing users to tune the library to the needs of a specific applications, the resulting performance can actually be worse compared to using the traditional socket implementation, as we show in Section \ref{evaluation}. Additionally, the default configuration is only suited to basic use cases (e.g. single threaded applications), requiring some time being spent on finding the right configuration for complex applications, using multiple threads and connections.

\textbf{SocksDirect}\cite{socksdirect} is a closed source library from Microsoft, written in C/C++. Like libvma, it works by preloading it to socket-based applications, redirecting socket traffic via a custom protocol based on RDMA. It also supports acceleration of intra-host communication via shared memory. It achieves low latencies and a high throughput by removing large parts of the synchronization and buffer management involved in traditional socket communication, while being fully compatible with linux sockets, even when process forking is involved.

We were able to get access to the source code from the authors and have successfully tested it with native applications, but so far we could not get the library working with Java applications. Additionally, SocksDirect uses the experimental verbs API, only available in the Mellanox OFED up to version 4.9\cite{verbs_exp}.

\subsection{Java libraries}
The \textbf{Sockets Direct Protocol) SDP}\cite{sdp} provided transparent offloading of socket traffic via RDMA, fully bypassing the kernel's network stack. It was part of the OFED package and introduce into the JDK starting with Java 7. However, support has officially ended and it has been removed from the OFED in version 3.5\cite{ofed35releasenotes}

\textbf{Java Sockets over RDMA (JSOR)}\cite{jsor} has been developed by IBM with the goal to offload all socket traffic of Java applications to high-speed NICs using RDMA. It is included in the IBM SDK up to version 8, requiring their proprietary J9 JVM. JSOR is not available in newer SDK versions and while the old SDK still receives security updates, applications using features not available in Java 8 cannot be used with JSOR.

While it has shown promising results in our benchmarks, there are known problems with connections getting stuck\cite{jsor-issue1} and exceptions\cite{jsor-issue2}. Additionally, we were not able to evaluate JSOR using a bidirectional connection with separate threads for sending and receiving. These problems and its reliance on on proprietary technology limit its usability, especially for modern applications.

\subsection{Application-specific solutions}
Other approaches aim at accelerating network performance of a specific application or framework. In 2014, a successful attempt at redesigning Spark's shuffle engine for RDMA usage has been made\cite{spark-rdma1} and refined in 2016\cite{spark-rdma2}. Similar solutions have been implemented for Apache Storm: In 2019, RJ-Netty has been proposed as a replacement for netty in Apache Storm\cite{rj-netty}, while in 2021 another approach at integrating RDMA into Storm, based on DiSNi\cite{disni} (formerly jVerbs\cite{jverbs}) has been implemented\cite{storm-rdma}.

While these solutions show, that the performance benefit for using high-speed networking hardware can be huge, they are specific to a single framework only and can not be used for general purpose network programming, like transparent acceleration libraries.
\section{Supporting netty in hadroNIO}
\label{hadronio-changes}

While hadroNIO has been working with applications directly using Java NIO, we encountered new challenges with netty-based applications. This section presents these challenges and their solutions, as well as changes in the design of hadroNIO. For a general overview of our architecture and the Java NIO API, as well as UCX, we refer to our original paper\cite{hadroNIO}.

\begin{figure}[h]
	\includegraphics[width=\linewidth]{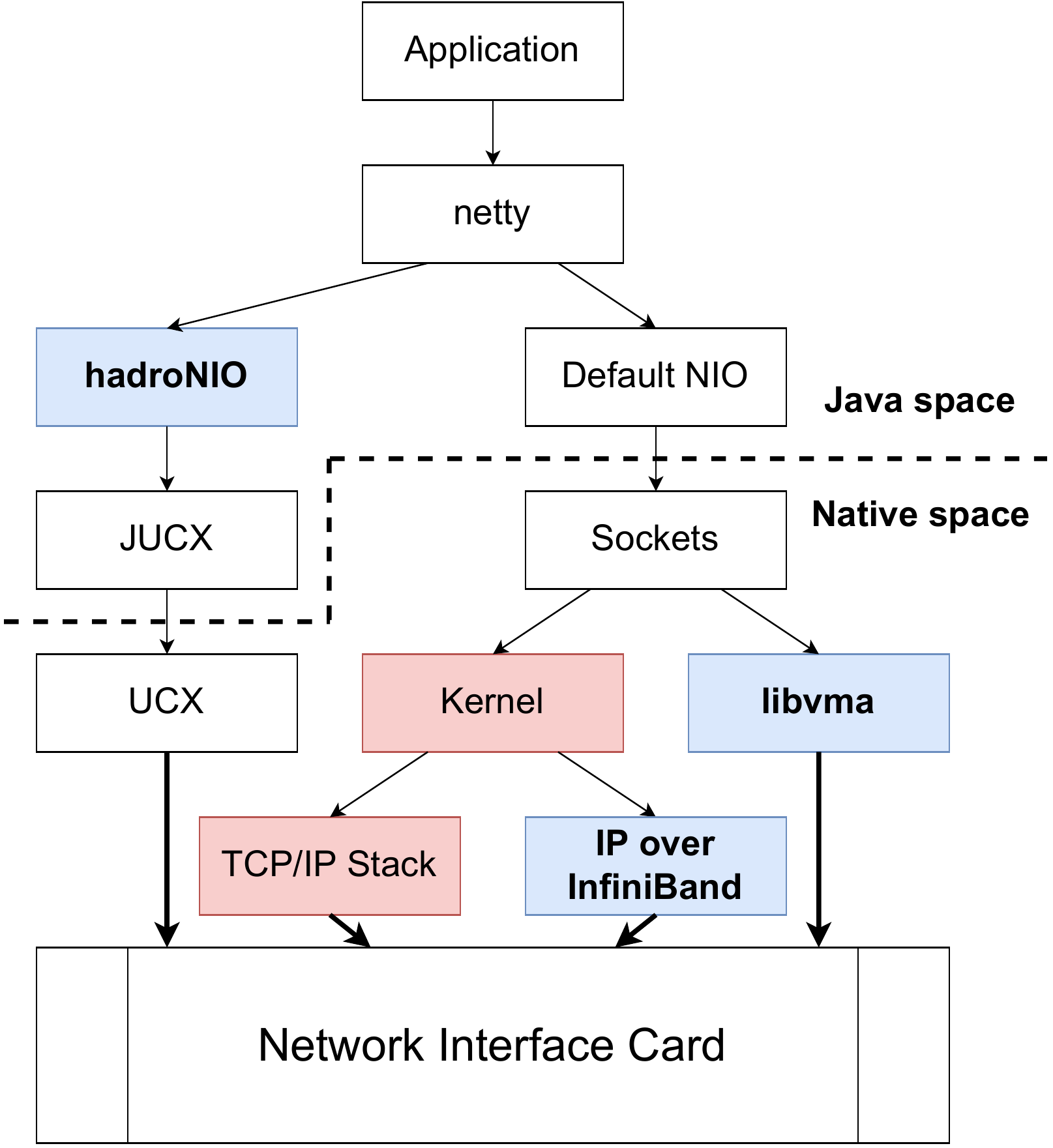}
	\caption{Application stack overview}
	\label{fig:fig1}
\end{figure}

The full application stack for hadroNIO, libvma, IP over InfiniBand and traditional sockets from netty to NIC is depicted by Fig. \ref{fig:fig1}.

\subsection{Providing a direct socket reference for netty}
A Java NIO \texttt{SocketChannel} provides access to its underlying socket via the \texttt{socket()} method. Since this would defeat the purpose of NIO, accessing a socket directly via a channel, is generally not done. However, netty keeps a reference to the socket of each channel to access its configuration (e.g. buffer size).

Since hadroNIO directly replaces the default NIO implementation, there is no underlying socket. In contrast to the aforementioned transparent acceleration solutions, we consciously chose to intercept traffic at the NIO level, instead of the socket level, since it fits well with the UCX API and has a more modern interface than traditional Java sockets. In its initial version, hadroNIO would throw an \texttt{UnsupportedOperationException} when \texttt{socket()} is called. However, for compatibility with netty, we needed to provide a workaround for the socket access, which we implemented in the form of two classes called \texttt{WrappingSocket} and \texttt{WrappingServerSocket}, extending the JDK's \texttt{Socket} and \texttt{ServerSocket} classes. They wrap an instance of a \texttt{SocketChannel}, or \texttt{ServerSocketChannel} respectively, and implement methods to access connection attributes, such as IP addresses and buffer sizes.



Once a connection has been terminated, the respective socket channel becomes readable, indicated by the \texttt{OP\_READ} flag. However, each attempt at actually reading data from the channel will return \texttt{-1}, signalling a closed connection. This behaviour was not implemented in earlier version of hadroNIO, since it only affects connection termination and our benchmarks would run without it. However, for full compatibility with the NIO specifications, we retrofitted it.


\subsection{UCX worker management}
UCX uses so called \emph{endpoints} to represent connections. However, these endpoints cannot send/receive data on their own. Instead UCX introduces the concept of \emph{workers}, which serve as an abstraction between endpoints and network resources (i.e. NICs). Each worker can be associated with multiple endpoints.

In the original design of hadroNIO, we used a single worker for all connections. However, since there is a limit on the maximum amount of connections a worker can handle, we refined our architecture to use multiple workers. Originally, we planned to use one worker per selector, which appeared as a natural fit, because a selector is used to query multiple channels, while a worker can progress multiple endpoints. However, NIO allows reassigning of channels to different selectors, which is not possible with UCX endpoints and workers. Ultimately, we settled on using a single worker per connection. This added complexity to our selector implementation, since it now has to poll multiple workers, but makes channels independent from selectors and allows reassignments.

\subsection{Supporting netty write aggregation}
\label{gathering-writes}
Java NIO offers two methods for sending data via a socket channel: One only takes a single buffer, while the other one is prescribed by the interface \texttt{GatheringByteChannel}\cite{gathering-channel}, thus capable of gathering write operations, accepting an array of buffers to send. Gathering writes are used heavily by netty (see chapter \ref{throughput-benchmark}) to bundle multiple send requests into a single method call, in order to achieve higher throughputs. However, in the initial hadroNIO version, we implemented the gathering write method by simply looping over all buffers, sending each one separately using the single buffer write method. While this implementation worked correctly, it did not offer any performance improvements, which is why we reimplemented it. Now, as many buffers as possible are merged into a single contiguous space inside hadroNIO's outgoing ring buffer, requiring only a single UCX write request to send. This massively improved throughput rates with netty-based applications.


\section{Benchmark architecture}
\label{benchmark-architecture}

To evaluate the performance of different acceleration solutions with netty-based applications, we designed and implemented two microbenchmarks, using netty for connection establishment and data exchange: One is focussed on throughput while the other implements a ping-pong pattern to measure round-trip times. The benchmarks are designed to work on two nodes of a cluster environment with one acting as a server and one acting as a client. Both support on or multiple connections between server and client and each connection is handled by a separate thread. Measurements are taken per connection, and a final result, taking all measurements into account, is calculated at the end.

\subsection{Connection setup}
The connection setup is similar for both benchmarks: On startup, the server sets up a server channel to listen for incoming connections. It then waits until a specified amount of connections has been established. Once the amount is reached, all threads start sending messages at the same time (throughput benchmark) or send a single message to kick off the ping-pong pattern (latency benchmark). Before the actual benchmark starts, a tenth of the operations are executed as warm up, without taking any measurements.

The client on the other side just needs to establish the specified amount of connections and wait for the server to start the benchmark.

\subsection{Throughput benchmark}
\label{throughput-benchmark}
Once all connections are set up, the server starts a separate thread for each connection, responsible only for sending messages through the respective channel. Once all warmup messages are sent, the thread waits for a synchronization message from the client, signalling that all messages have been received successfully. Each thread then needs to pass a barrier, ensuring that all threads start the benchmark at the same time. After all benchmark messages have been sent, the client once again sends a signal to server, finishing the benchmark. Times are measured once after the warmup barrier has been passed and after the second signal from the client has been received. allowing us to calculate the average data and operation throughput rates.

When sending a buffer via netty, it is not transmitted directly, but first stored in an instance of a class called \texttt{ChannelOutboundBuffer}\cite{netty-outbound}, which accumulates outgoing write requests. To make sure, that data is actually transmitted, applications need to manually flush the respective channel. The data, contained in a buffer, is not copied, but only references to all outgoing buffers stored. Once netty is requested to perform a flush operation, all buffers are send with a minimal amount of write operations, using the gathering write method described in chapter \ref{gathering-writes}. This aggregation strategy allows netty to reach high throughputs without requiring any buffer copies. Our throughput benchmark can be configured to use a specific interval (e.g. every 64 buffers) for flushing a channel, allowing us to analyse performance with different amounts of aggregated buffers.

\subsection{Latency benchmark}
The latency benchmark does not start threads on its own, but makes use of netty's worker threads. Each time data is received, a worker thread invokes a method in the respective handler (instance of \texttt{ChannelInboundHandlerAdapter}\cite{netty-inbound}) and once our handler implementation has received a full message, it issues a write request, following a ping-pong pattern. We configure netty to start as many worker threads, as there are connections, with each thread opening its own selector and connections being assigned to these selectors in a round-robin fashion. This ensures, that each connection has its own thread, responsible only for handling requests on that specific connection. Times are measured before each send call and after each received message, allowing us to gather the round-trip latencies of all operations.
\section{Evaluation}
\label{evaluation} 

This section presents and discusses the evaluation results, comparing default netty performance using sockets via Ethernet versus accelerating netty with hadroNIO and libvma using 100 GBit/s high-speed NICs.

\subsection{Evaluation setup}
We used the microbenchmarks described in chapter \ref{benchmark-architecture} for evaluating messaging performance with netty, regarding throughput, as well ass round-trip latency in two different cluster environments. To test the scalability of each solution, we increased the connection count step-wise from 1 to 16.

Our benchmark environment consisted of two identical bare-metal nodes, provided by the Oracle Cloud Infrastructure, using the \emph{HPC Cluster} Terraform stack\cite{oci-stack}:
\begin{figure}[h]
	\begin{tabularx}{\linewidth}{|r|X|}
		\hline
		\textbf{CPU} & 2x Intel(R) Xeon(R) Gold 6154 CPU (18 Cores/36 Threads @3.00 GHz) \\
		\hline
		\textbf{RAM} & 384 GB DDR4 @2933 MHz \\
		\hline
		\textbf{NIC} & Mellanox Technologies MT28800 Family [ConnectX-5] (100 GBit/s) Ethernet \\
		\hline
		\textbf{OS} & Oracle Linux 7.9 with Linux kernel 3.10.0-1160 \\
		\hline
		\textbf{OFED} & MLNX 5.3-1.0.0.1 \\
		\hline
		\textbf{Java} & OpenJDK 17.0.2 \\
		\hline
		\textbf{UCX} & 1.12.1 \\
		\hline
		\textbf{hadroNIO} & 0.3.2 \\
		\hline
		\textbf{libvma} & 9.5.0 \\
		\hline
	\end{tabularx}
	\caption{Hardware specification of the OCI systems.}
	\label{fig:fig2}
\end{figure}

We evaluated throughput and latency with small (16 byte) mid-sized (1 KiB) and large (64 KiB) messages. For evaluating throughput, we sent 100 million messages per benchmark run, while 10 million round-trip operations were executed during each latency benchmark run. For the large buffers, we used 10 million and 1 million messages respectively and evaluated with up to 12 connections, to avoid unnecessary long running benchmarks. The amount of connections is always depicted by the y-axis, while the x-axis shows the data throughput in MB/s or GB/s when looking at throughput results, and the round-trip time in \textmu s when evaluating latency. Each benchmark run was executed five times and the graph depicts the average values, while the error bars show the standard deviation.

\subsection{Configuration and Optimizations}
Each of the OCI nodes had two CPUs with 18 cores and 36 threads each at its disposal. To optimize performance, we used the tool \emph{numactl} to bind the JVM process to the processor, that the network card is connected to. Since a single CPU has 18 cores, it should not be overwhelmed by 16 connections at once.
The ConnectX-5 NICs were configured to use Ethernet as the link layer protocol, making them fully compatible with traditional sockets.

To improve performance regarding the throughput benchmarks, we did not flush the channels after each written message, but gave netty the chance to gather multiple message and send them at once. For small messages, we flushed each time 64 messages were written and for mid-sized and large messages, we used intervals of 16 and 4 messages respectively.

To work correctly, libvma needs to either be executed by the root user or with the CAP\_NET\_RAW privilege. We tried granting CAP\_NET\_RAW as described in libvma's README file\cite{libvma-readme}, but could not get it to offload traffic. Fortunately, running as the root user worked in the OCI environment.

Additionally, we set the amount of hugepages to 800 and shmmax to 1000000000, as recommended\cite{libvma-readme}. Furthermore, libvma exposes a lot of configuration parameters, settable via environment variables. As endorsed by the libvma wiki, we set VMA\_RING\_ALLOCATION\_LOGIC\_RX and VMA\_RING\_ALLOCATION\_LOGIC\_TX to 20, which should improve multithreading performance\cite{libvma-parameters}. We also needed to increase the amount of receive buffers via VMA\_RX\_BUFS to 800000, otherwise the benchmark would sometimes not finish with 12 or more connections, because libvma ran out of buffers. For the round-trip measurements, we set VMA\_SPEC to \emph{latency}.

For hadroNIO, we used the default configuration with 8 MiB large ring buffers and a slice length of 64 KiB.

\subsection {Small messages (16 byte)}
\begin{figure}[h]
	\includegraphics[width=\linewidth]{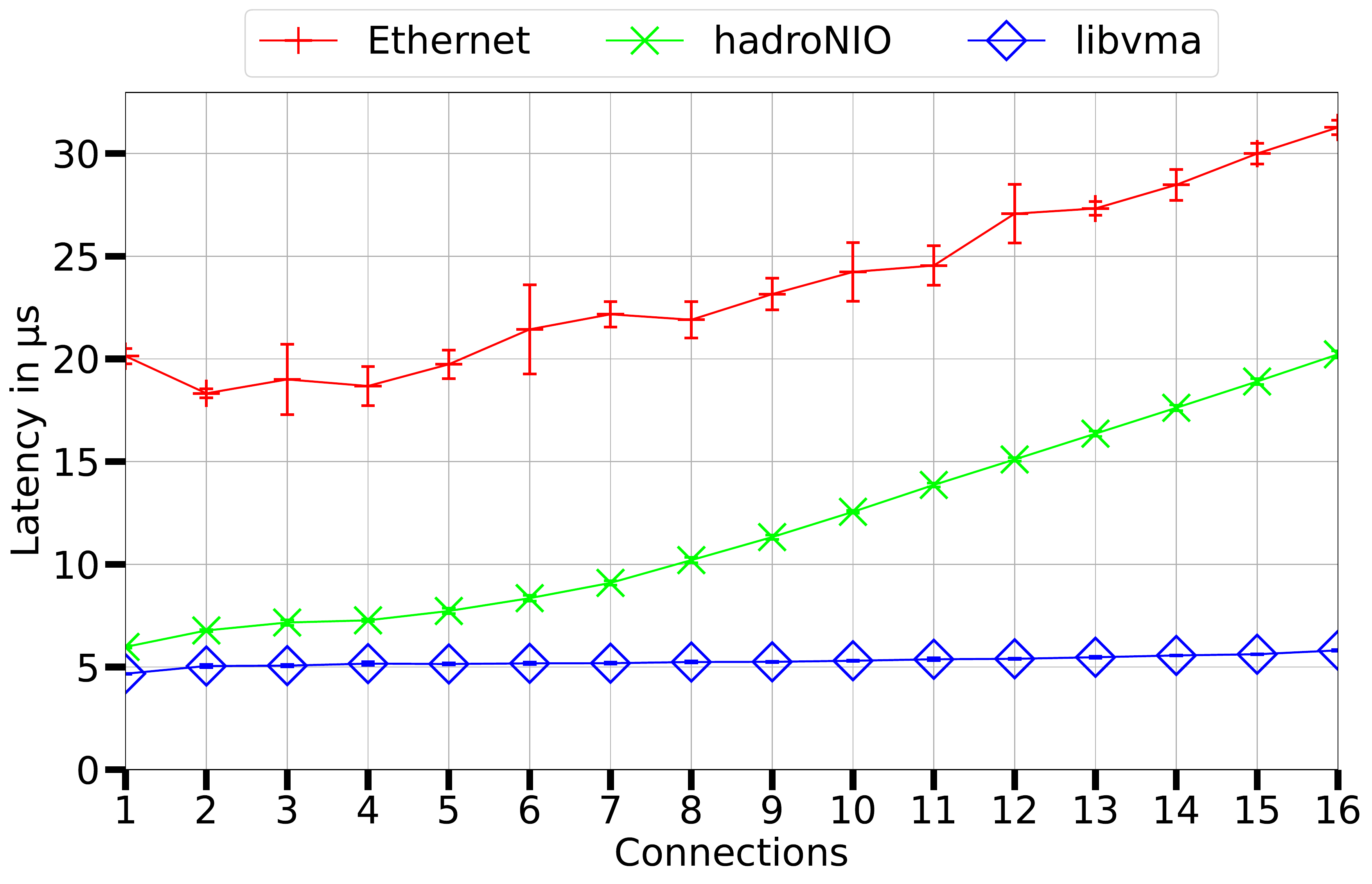}
	\caption{Average round-trip times with 16-byte messages}
	\label{fig:fig3}
\end{figure}
Starting with 16-byte messages, Fig. \ref{fig:fig3} shows the round-trip times for traditional Ethernet, hadroNIO and libvma. As can be seen, libvma offers the best latency, with almost no overhead being generated by using multiple connections. Starting with 4.7 \textmu s using a single connection, it still manages to yield round-trip times of 5.8 \textmu s with 16 parallel connections.

While hadroNIO offers similarly low latencies with few connections, starting with 6 \textmu s, it breaks the 10 \textmu s mark using 8 connections. From there on, each additional connection adds around 1 \textmu s of latency.

However, both acceleration solutions offer a substantial performance improvement over plain Ethernet, which starts at 20 \textmu s using a single connection. Curiously, round-trip times fall to around 18 \textmu s for 2-4 connections but constantly rise starting with 5 connections.

\begin{figure}[h]
	\includegraphics[width=\linewidth]{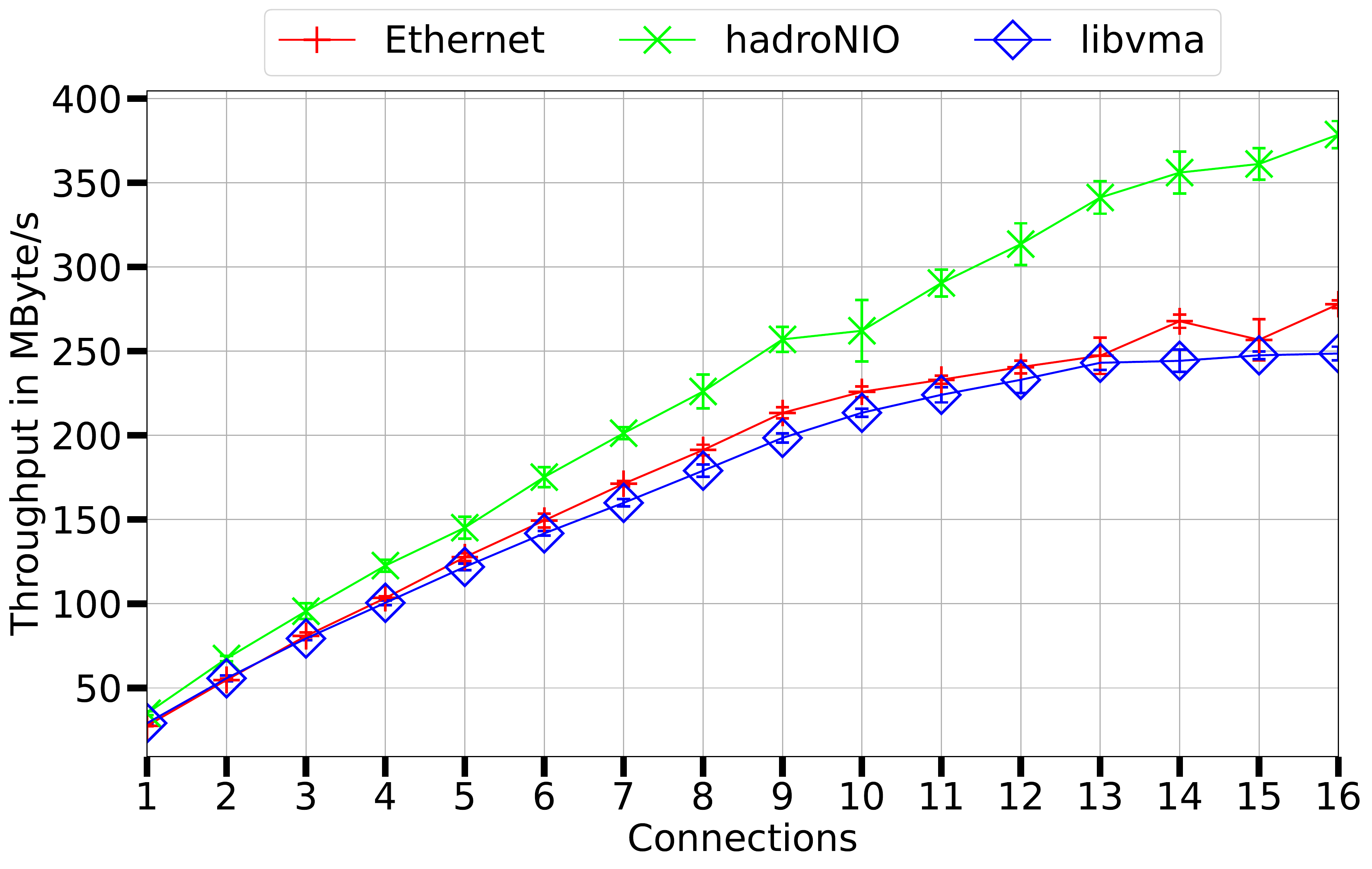}
	\caption{Average throughput with 16-byte messages}
	\label{fig:fig4}
\end{figure}
The throughput values, depicted by Fig. \ref{fig:fig4}, paint a different picture. When using only one connection, all three solutions offer similar performance between 28 and 35 MB/s, with hadroNIO having a slight advantage. However, with a rising connection count, the gap between hadroNIO and Ethernet/libvma grows larger, with libvma even offering slightly lower throughput values than plain Ethernet. Starting with 13 connections, libvma almost completety stops scaling, reaching around 250 MB/s, while hadroNIO scales further to 380 MB/s using 16 connections.

While libvma offers the smallest round-trip times with small messages, its throughput rates are slower than using Ethernet, whereas hadroNIO scales much better than the other candidates in our throughput benchmark.

\subsection{Mid-sized messages (1 KiB)}
\begin{figure}[h]
	\includegraphics[width=\linewidth]{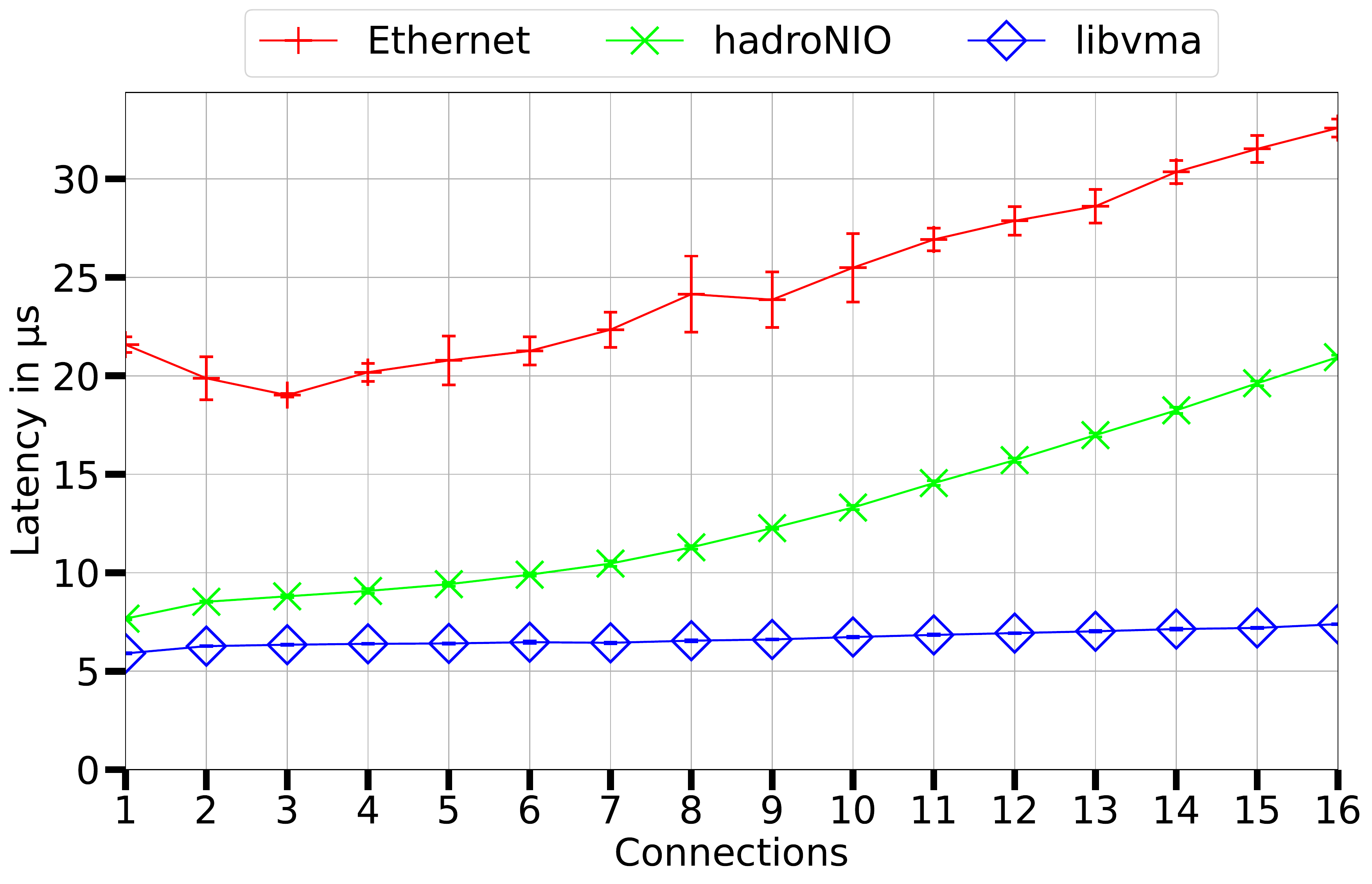}
	\caption{Average round-trip times with 1 KiB messages}
	\label{fig:fig5}
\end{figure}
Looking at the round-trip times for 1 KiB payloads (see Fig. \ref{fig:fig5}), the three solutions perform almost the same compared to the 16-byte results, apart from an offset being added to all latencies. Again, libvma scales almost perfectly, starting with 5.9 \textmu s for a single connection and only rising to 7.4 \textmu s using 16 connections, while hadroNIO starts with 7.6 \textmu s, with slowly rising latencies up to 10.5 \textmu s using 7 connections and linear increasing values from there on.

\begin{figure}[h]
\includegraphics[width=\linewidth]{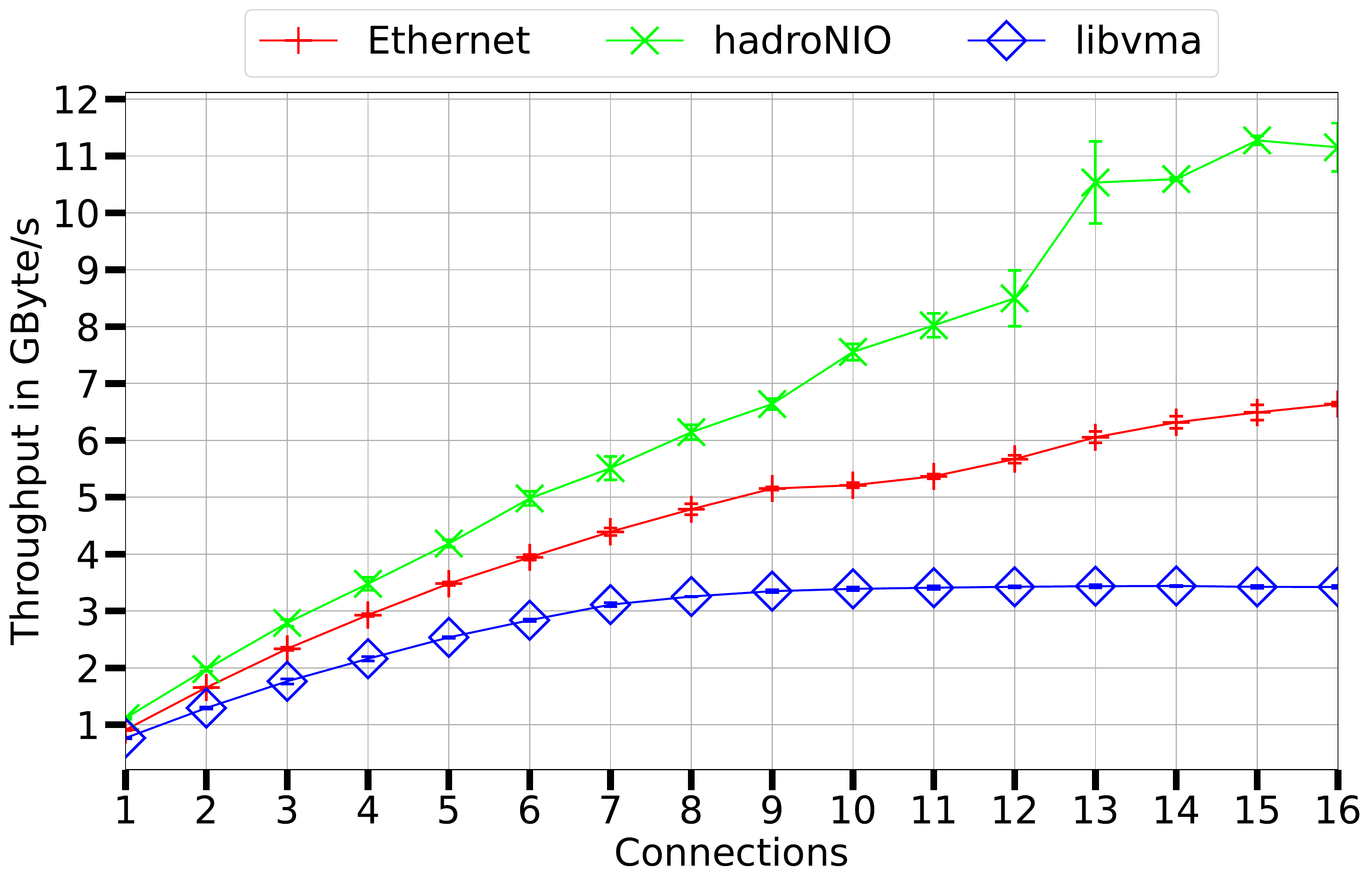}
\caption{Average throughput with 1 KiB messages}
\label{fig:fig6}
\end{figure}
The throughput values, shown in Fig. \ref{fig:fig6}, demonstrate that hadroNIO again scales well with an increasing amount of connections, reaching more than 11 GB/s at the end, thus almost saturating the 100 GBit/s hardware. On the other side, libvma scales much slower and reaches its top speed of just 3.4 GB/s with 10 parallel connections. The same throughput can be achieved using hadroNIO with only 4 connections and even using no acceleration solution at all is substantially faster, surpassing libvma's maximum throughput using 5 threads and reaching around 6.6 GB/s with 16 threads.

To conclude the evaluation of mid-sized messages, libvma continues to offer the best performance with regards to round-trip times, but comparing the results of the throughput benchmark, it falls behind hadroNIO and even plain Ethernet by far.

\subsection{Large messages (64 KiB)}
\begin{figure}[h]
	\includegraphics[width=\linewidth]{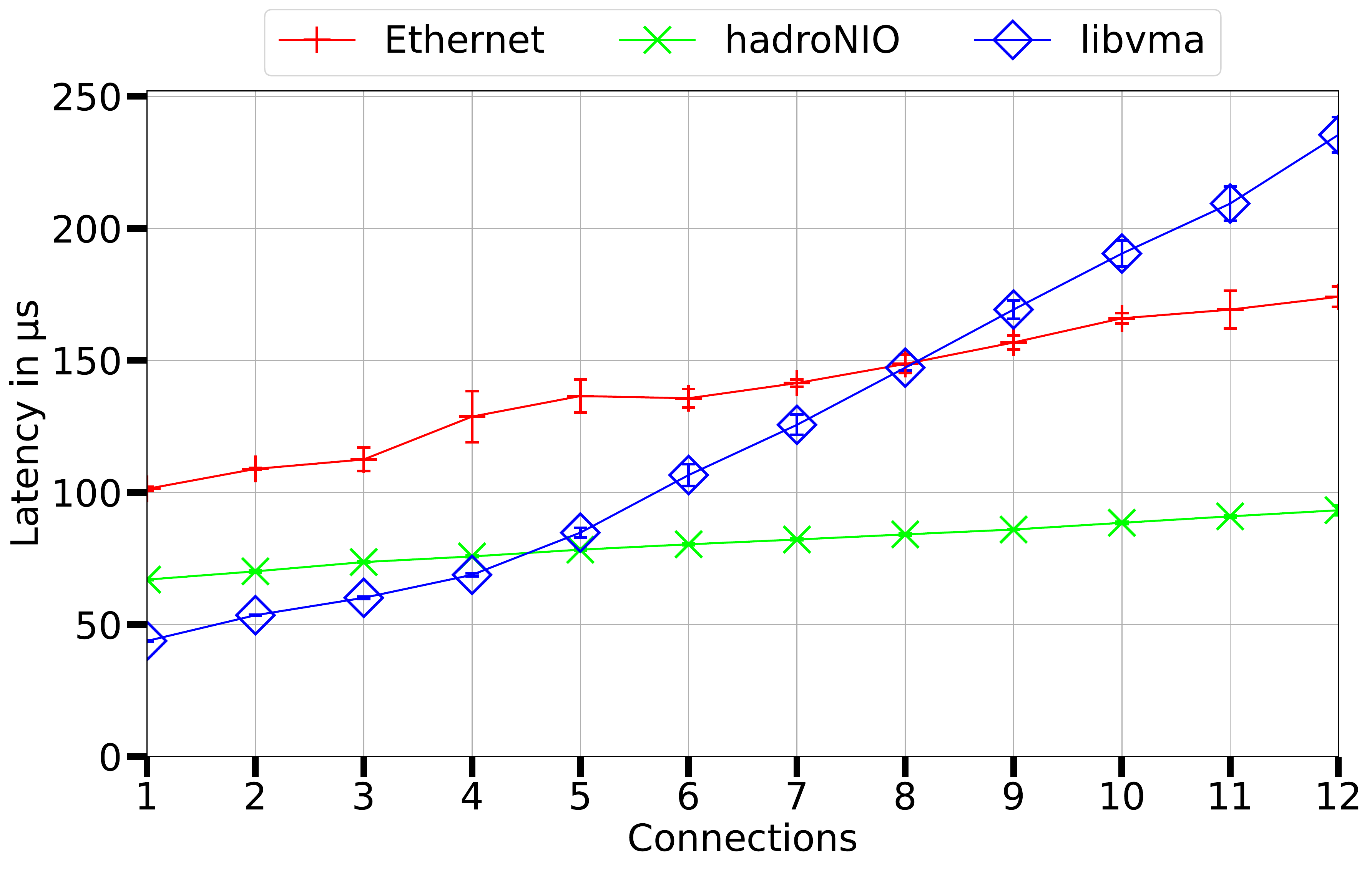}
	\caption{Average round-trip times with 64 KiB messages}
	\label{fig:fig7}
\end{figure}
Continuing with large 64 KiB payloads, the latency results, depicted by Fig. \ref{fig:fig7}, differ from the previous ones. While libvma yields the lowest round-trip times for up to 4 connections (44-69 \textmu s), values increase faster from there on, rising by around 20-25 \textmu s per additional connection. Starting with 9 parallel connections, libvma performs worse than plain Ethernet and the gap grows further with an increasing thread count. While hadroNIO yields higher latencies than libvma for 1-4 connections (67-76 \textmu s), it offers the best performance using 5 or more parallel connections, reaching round-trip times of only 94 \textmu s using 12 threads, while libvma is 2.5 times slower with around 235 \textmu s.

\begin{figure}[h]
\includegraphics[width=\linewidth]{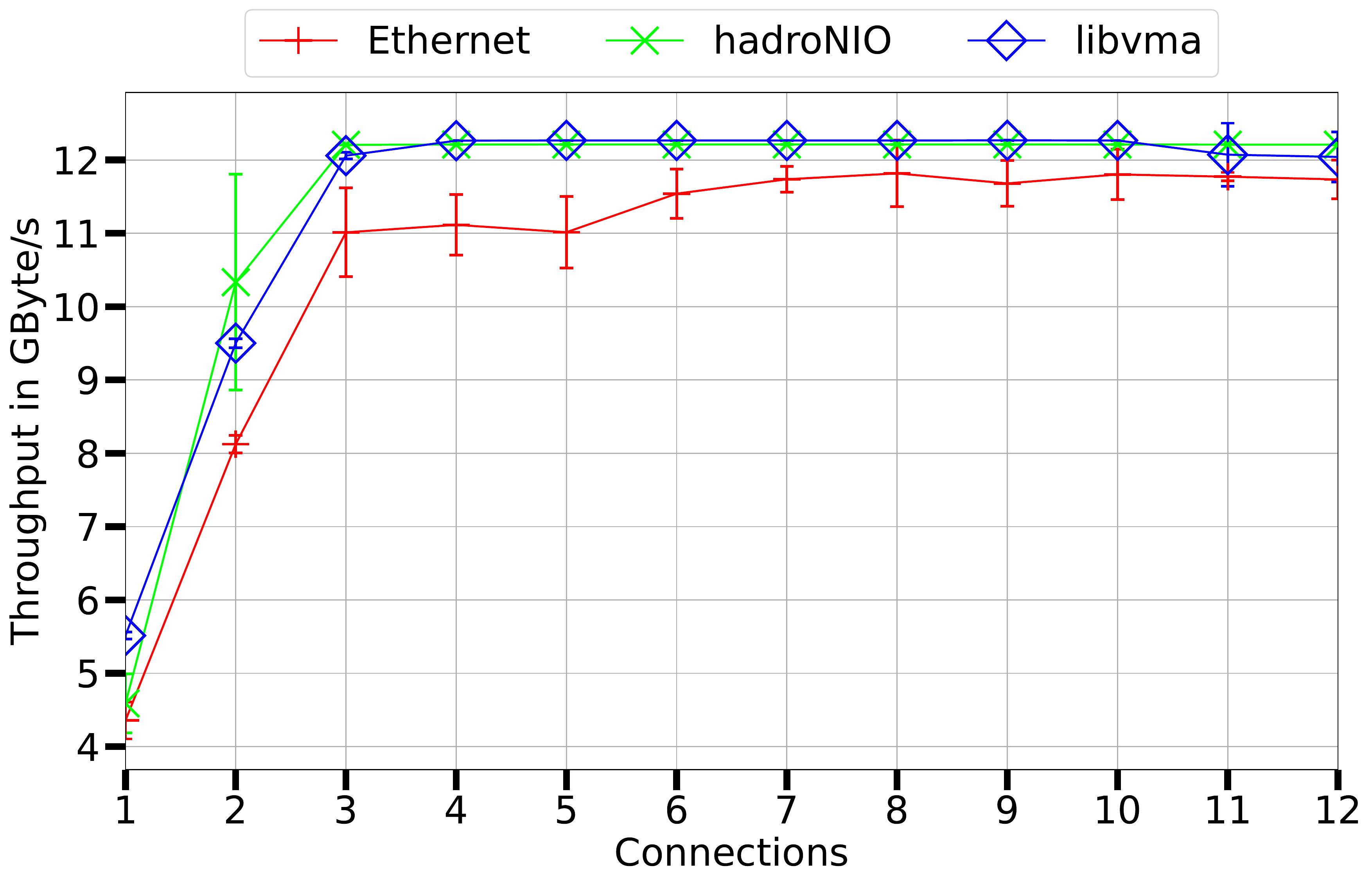}
\caption{Average throughput with 64 KiB messages}
\label{fig:fig8}
\end{figure}
We close the evaluation, by looking at the throughput values using 64 KiB messages. Both acceleration solutions offer similar performance, managing to saturate the NIC with more than 12 GB/s using 3 or more connections. For a single connection, libvma is faster with 5.5 GB/s versus 4.6 GB/s, but with 11 and 12 connections, libvma becomes somewhat unstable and falls slightly behind hadroNIO. Using plain Ethernet offers acceptable performance, but  12 GB/s cannot be reached and the results are not stable, with standard deviations sometimes as high as 1 GB/s.

Concluding the large payload results, both libvma and hadroNIO are able to saturate the hardware, but regarding round-trip times, it depends on the amount of connections, which solution performs best.
 \section{Conclustions \& Future Work}
 \label{conclusion}
In this paper, we presented hadroNIO extensions to support netty and compared the performance of netty based on hadroNIO versus libvma and traditional sockets over Ethernet using two microbenchmarks, for evaluating round-trip times and throughput. Our results show, that hadroNIO offers a substantial performance improvement over Ethernet on the same NIC, without needing elevated privileges or complex configurations. All results were achieved using hadroNIO's default configuration values. While libvma offers the lowest latency with small and mid-sized messages, preloading it to a netty-based application can actually worsen performance and it may not be usable in every environment due to it being dependent on CAP\_NET\_RAW or root privileges..

Future work includes evaluating hadroNIO with large netty-based applications and frameworks, such as Apache Cassandra and gRPC. We also aim to improve our selector implementation, by leveraging epoll, since it is currently based on busy polling. Additionally, we are working on integrating Infinileap , a UCX binding for Java, based on the experimental Foreign Function \& Memory API (Project Panama)\cite{panama}, into hadroNIO to see if the overhead introduced by JNI calls can be alleviated. Furthermore, we want to evaluate hadroNIO with GraalVM\cite{graalvm}, offering low-cost interoperability between Java and native code.
\section{Acknowledgment}
\ifblind
This section has been omitted for double blind reviews.
\else
We thank Oracle for their sponsorship in the context of this work.\\
This work was supported in part by Oracle Cloud credits and related resources provided by the Oracle for Research program.
\fi

\bibliography{paper}
\bibliographystyle{abbrv}


\end{document}